
\parskip=10pt
\parindent=0pt

\def\day{30/6/97}

\def\ref#1{[#1]}
\def\pa{\partial}
\def\~#1{{\widetilde #1}}
\def\({\left(}
\def\){\right)}
\def\[{\left[}
\def\]{\right]}
\def\ker{{\rm Ker}}
\def\A{{\cal A}}
\def\B{{\cal B}}

\def\la{\lambda}
\def\a{\alpha}
\def\b{\beta}

{\nopagenumbers
{\parskip=0pt
~ \vskip 2 truecm

\centerline{\bf Normal Forms, Symmetry,}
\medskip
\centerline{\bf and Linearization of Dynamical Systems.}
\footnote{}{{\tt Version of \day}  }

\vskip 4 truecm
{ Dario Bambusi},

{{\it Dipartimento di Matematica,
Universit\`a di Milano, via Saldini 50,
20133 Milano (Italy)};

{\tt bambusi@vmimat.mat.unimi.it}}
\smallskip
{ Giampaolo Cicogna},

{{\it Dipartimento di Fisica, Universit\`a di Pisa,
P.zza Torricelli 2, 56126 Pisa (Italy)};

{\tt cicogna@ipifidpt.difi.unipi.it}}
\smallskip
{ Giuseppe Gaeta},

{{\it Department of Mathematics, Loughborough University,
Loughborough LE11 3TU (G.B.)};

{\tt g.gaeta@lboro.ac.uk} }
\smallskip
{ Giuseppe Marmo},

{{\it Dipartimento di Scienze Fisiche,
Universit\`a di Napoli,} and {\it  INFN, Sezione di Napoli,
Mostra d'Oltremare, 80123 Napoli (Italy)}};

{\tt gimarmo@napoli.infn.it }

\vskip 2 truecm
}

{\bf Summary.} {We discuss how the presence of a suitable
symmetry can guarantee the perturbative linearizability of a
dynamical system -- or a parameter dependent family -- via
the Poincar\'e Normal Form approach. We discuss this at first
formally, and later pay attention to the convergence of the
linearizing procedure. We also discuss some generalizations of
our main result.}

\vfill\eject  }  \pageno=1

~
\vfill
{\bf Introduction.}
\bigskip

It is well known that the same procedure -- based on
formal series of polynomial changes of coordinates --
devised by Poincar\'e \ref{1-4} to integrate linearizable
dynamical systems\footnote{$^1$}{By dynamical system, we
will mean a polynomial ODE in $R^n$, or equivalently a
polynomial vector field on $R^n$. Similarly, by vector field
we will mean an analytic one.} in the neighbourhood of a
fixed point, can also be used to normalize, again in the
neighbourhood of a fixed point, non-linearizable dynamical
systems, i.e. system whose linearization at the fixed point
present resonances.

This fact suggests that the Poincar\'e procedure does not
take full advantage of the peculiar nature of (locally)
linearizable system, so that it is not impossible to
obtain, in this specific case, some improvement over the
general theory of (Poincar\'e-Dulac) Normal Forms.

It was recently proposed \ref{5} that the linearizability of
a dynamical system can be analyzed, and asserted under
certain condition, {\it non-perturbatively} by considering the
{\it symmetries} which are associated with the linearity of
the system in suitable coordinates.

Here we want to show how these symmetry properties come
into play -- to ensure linearizability of the system (by
means of formal, or possibly convergent, series of
near-identity changes of coordinates) -- in the framework
of the {\it perturbative} theory, i.e. in the theory of
Poincar\'e-Dulac Normal Forms \ref{1,2}. Thus,
although we investigate the same kind of question as in
\ref{5}, we operate in quite a different framework, and
we can make little use of results obtained  there,
as will be clear from the following; we discuss the relation
between \ref{5} and the present work, i.e. between the global
and the perturbative approach, in the appendix.

In order to discuss the linearizability of a system in the
perturbative approach, we have naturally to consider the case of
Normal Forms in the presence of symmetry. More
specifically, it turns out that we have to consider Normal
Forms in the presence of {\bf nonlinear} symmetry: indeed,
the symmetries associated with the linearizable nature of
the system may be linear only in the coordinates in which the
dynamical system is indeed linear, i.e. when the problem is
already solved; moreover such coordinates could be defined only
locally in a neighbourhood of the origin.
Thus, rather than employing the classical
theory of Normal Forms in the presence of linear symmetries
\ref{2,6-8}, we will use recent results \ref{9,10}, which
deal with the general (i.e. nonlinear) case.

Needless to say, systems which can be linearized -- albeit only
locally -- are highly nongeneric, and correspond to ones that
can be exactly solved locally. This fact shows at the same time the
limitation for applications of our result, as it cannot deal with
generic systems, and its interest, as it deals with systems which are
special but also specially interesting, both in themselves and as
starting points for a perturbative analysis of more general ones.

It should be mentioned that the connection between
``suitable'' symmetries and linearity of the Normal Form
was actually already remarked -- albeit {\it en passant} --
in \ref{9} (see remark 4 there); we want to discuss this in
more detail for two reasons: on one side, for its practical
relevance; and on the other because it shows how
consideration of nonlinear symmetries in Normal Form theory
really improves the results that can be obtained by
considering only linear symmetries.

\vfill\eject
{\bf 1. The main result.}
\bigskip

Let us consider a vector field $X$, which in the
coordinates $(y_1 , ... , y_n )$ on $R^n$ is simply given by
$$ X \ = \ \sum_{i=1}^n \ A_{ij} \ y_j \ {\pa ~ \over \pa
y_i } \ , \eqno(1) $$
where $A$ is a real $(n \times n)$ matrix. This vector
field does obviously commute with all the vector fields
$$ Y_{(k)} \ = \ \sum_{i=1}^n \ [A^k]_{ij} \ y_j \ {\pa ~
\over \pa y_i } \eqno(2) $$
associated to the (non-negative) integer powers of the
matrix $A$. In particular, whatever the matrix $A$, $X$
does commute with the vector field $S \equiv Y_{(0)}$ which
generates the dilations in $R^n$,
$$ S \ = \ \sum_{i=1}^n \ y_i \ {\pa ~ \over \pa y_i } \ ,
\eqno(3) $$
and it is easy to see that, conversely, the only vector
fields commuting with $S$ are the linear ones. This
observation is, of course, completely trivial;
nevertheless, it is extremely useful in studying
linearizability of nonlinear systems.

Let us now consider a (formal)
near-identity [this means that $(D \phi
)(0) = I$] nonlinear change of coordinates,
$$ y_i = \phi_i (x , ... , x_n ) \ . \eqno(4) $$
Under this, the linear dynamical system
$$ {\dot y_i} \ = \ A_{ij} \ y_j \eqno(5) $$
is changed in general into a nonlinear system
$$ {\dot x^i} \ = \ f^i (x) \ ; \eqno(6) $$
it is immediate to see that the functions $f^i (x)$ are
given by
$$ f^i (x) \ = \ \[ J^{-1} (x) \]_{ij} \ A_{jk} \ \phi_k
(x) \ , \eqno(7) $$
where $J$ is the jacobian of the coordinate transformation,
$$ J_{ij} \ = \ {\pa y_i \over \pa x_j } \ \equiv \ {\pa
\phi_i \over \pa x_j} \ . \eqno(8) $$
Notice that, as we assumed (4) to be a near-identity
transformation, the inverse of the jacobian exists, at
least in some neighbourhood of the origin, so
that (7) makes sense.
Again by the near-identity of (4), we are guaranteed that
$(D f)(0) = A$.

{\tt Remark 1.} It should be noted that the theory can also be
formulated in terms of Lie transformations \ref{11,12}: in this
case the coordinate transformation correspond to the time-one
flow of an analytic vector field, and a number of technical
problems -- in particular, concerning inverse
transformations -- are automatically taken care of. Here we
stick to the usual setting for the sake of simplicity. $\odot$

In the new coordinates, $X$ is expressed as
$$ X \ = \ \sum_{i=1}^n \ f^i (x) \ {\pa ~ \over \pa x_i}
\equiv f^i\pa_i \eqno(9) $$
Similarly, $S$ is now expressed as
$$ S \ = \ \sum_{i=1}^n \ p^i (x) \ {\pa ~ \over \pa x_i}
\equiv p^i\pa_i \ , \eqno(10) $$
where the $p_i$ are nonlinear functions given explicitely by
$$ p^i (x) = \[ J^{-1} (x) \]_{ij} \phi_j (x) \ . \eqno(11)
$$

The geometrical relations between geometrical objects (in
particular, vector fields) do not depend, however, on the
choice of coordinates; thus, $[X,S] = 0$ continue to hold.
We recall that, if two vector fields $X=f^i\pa_i$ and $Y=g^i\pa_i$
satisfy $[X,Y]=0$, this means that, in terms of the components of
the vector fields,
$$ \{ f , g \}^i \equiv \( f^j \cdot \pa_j \)  g^i  - \( g^j
\cdot \pa_j \)  f^i = 0 \quad \quad \forall i = 1,..., n \ \ .
\eqno(12) $$

{\tt Remark 2.}
  Let us remark, however, that a vector field $X$ may be linear in two
  different coordinate systems even if they are connected by a {\it
  nonlinear} transformation: consider, e.g., the 1-dimensional harmonic
  oscillator
  $$ \eqalign{\dot q \ =& \ ~~ p  \cr \dot p \ =& \ - q \ , \cr} $$
  and a (nonlinear) transformation
  $$ \eqalign{ Q \ =& \ f(q^2+p^2)\ q  \cr  P \ =& \
  f(q^2+p^2)\ p \ , \cr} $$
  where $f$ is a smooth function such that $f(0)=1$. In the new
  coordinate system $Q,P$, one gets
  $$ \eqalign{\dot Q \ =& \ ~~ P  \cr  \dot P \ =& \ - Q \ , \cr} $$
  and the dynamics is linear. Therefore, in this case, the dilation fields
  $$ S = q{\pa\over{\pa q}} + p{\pa\over{\pa p}}   \qquad {\rm and} \qquad
    S' = Q{\pa\over{\pa Q}} + P{\pa\over{\pa P}} $$
  provide two {\it different} linear structures which linearize the
  vector field along with the respective symmetries, but are not taken
  into each other by the same transformation (indeed, $S = [1 + 2 (q^2 +
  p^2) f'/f ] S'$).
$\odot$

Suppose now that we have to study the {\bf nonlinear}
system (6), without knowing about (4), and in particular we
want to know if it is linearizable. It is quite easy to
produce examples in which, as e.g. in the examples we consider
below, the Poincar\'e-Dulac theory fails to recognize
immediately the intrinsically linear nature of
the system.

{\tt Remark 3.} It should maybe be specified that by this, we
mean that the Normal Form unfolding allows for nonlinear terms;
obviously, if one was going to actually perform the
normalization -- rather than just determining the general Normal
Form unfolding corresponding to the linear part -- there would
be a series of ``miracoulous'' cancellations, so that the
coefficients of the nonlinear resonant terms would vanish.
$\odot$

If we study the symmetry properties of (6), i.e. the vector fields  $Y =
g^i (x) \pa_i$  for which (12) is satisfied,
we know that {\it if (6) is linearizable, then (12) admits at least one
solution with $g=p$ given by (11)}. Notice that, in particular, this means
that -- also in the $x$ coordinates -- the linearization of $Y$ is just
given by $(DY)(0) = I$; this is again a consequence of the fact
(4) is near-identity. (Here and in the following we use the
short notation $(DY)(0)=M$, with $Y=g^i \pa_i$ a vector field
and $M$ a matrix, to mean that $(Dg)(0) = M$).

It is interesting to remark, and this constitutes our first result,
that the converse is also true; i.e., we have that:

{\bf Theorem 1.} {\it Let $X = f^i (x) \pa_i$ be a vector
field in $R^n$. If the equation $[X,Y]=0$ admits a solution
for which $(DY)(0) = I$, where $Y = g^i (x) \pa_i$, then $X$
is linearizable by a formal series of near-identity change of
coordinates. Moreover, there is a formal series of near-identity
change of coordinates which linearizes $X$
and which does also linearize $Y$, transforming it in the dilation
field $S$, and this happens for any such solution.}

The reason for this lies in a very simple consequence of a theorem given
in \ref{9,10}, which we report here for completeness, in a form suitable
for our present purpose (a stronger form of this theorem -- not relevant
to the present discussion -- also exists, see \ref{9,13}); in order to
state this we have to introduce some terminology.

We will refer to the classical S+N decomposition of a matrix:
this is the unique decomposition of a matrix $M$ into a
semisimple and a normal part, called respectively $M_s$ and
$M_n$, which moreover satisfy $\[ M_s , M_n \] = 0$ (see
e.g. \ref{14}).

We will also refer to (Semisimple) Joint Normal Forms: denote by
$\A , \B , \A_s , \B_s$ the homological operators \ref{1}
associated to $A , B , A_s , B_s$ (with the
notation introduced in (12), $\A = \{ Ax , . \}$, and so on).
We say that
$$ X = \~f^i (y) \( \pa / \pa y_i \) \quad , \quad Y = \~g^i
(y) \( \pa / \pa y_i \) \ , \eqno(13) $$
where
$$ \~f (y) = Ay + \~F (y)  \quad , \quad \~g (y) = B (y) +
\~G (y) \ , \eqno(14) $$
are in Semisimple Joint Normal Form if both $\~F$ and $\~G$ are
in $ \ker (\A_s ) \cap \ker (\B_s ) $, and that they are in
Joint Normal Form if both $\~F$ and $\~G$ are in $ \ker (\A ) \cap
\ker (\B ) $.

{\tt Remark 4.} We recall \ref{1,2,15} that
for $A$ normal, $\ker (\A ) = \ker (\A^+ )$; we also recall
that $\ker (\A )$ and $\ker (\A^+ )$ are contained in $\ker
(\A_s )$. $\odot$

With this notation, we can state the

{\bf Proposition.} \ref{9,10}
{\it Let the polynomial vector fields $X$ and $Y$ in $R^n$,
expressed in the $x$ coordinates as
$$ X = f^i (x) \pa_i \quad , \quad Y = g^i (x) \pa_i \ ,
\eqno(15) $$
commute, i.e. $[X,Y]=0$. Let the linearization
of $X$ and $Y$ at $x=0$ be given, respectively, by $A =
(DX)(0)$ and $B=(DY)(0)$, and let the matrices $A$ and $B$
have S+N decompositions $A=A_s + A_n$, $B= B_s + B_n $; let
$F$, $G$ be the nonlinear parts of $f$, $g$, so that $ f =
Ax + F$, $g = Bx + G$. Then, by a formal series of
near-identity (Poincar\'e) transformations, it is possible
to reduce $f,g$ to {\rm Joint Semisimple Normal Form}; if $A$
and $B$ are normal matrices, then it is possible to reduce
$f,g$ to {\rm Joint Normal Form}.}

{\tt Proof of Theorem 1.} Under the hypotheses of the above
Theorem, $B=I$, so that $B_s = B = I$; in this case in
particular  $\ker (\B ) = \ker (\B^+ ) = \ker (\B_s ) =
{\cal K} $, and, indipendently of $A$, we can,
due to the proposition, transform $f$ to $\~f$ with $\~F
\in {\cal K}$. It is a general result that ${\cal K} =
\ker (\B^+ )$ consists of vector polynomials which
are {\it resonant with $B$}; these are characterized as
follows \ref{1}. Let $\la_1 , ... , \la_n$ be the
eigenvalues of $B$; then ${\cal K}$ is spanned by vectors
$v= (v_1 , ... , v_n)$ which have all components equal to
zero except for $v_r$, which is given by $v_r (x) =
x^{m_1} ... x_n^{m_n}$, where the $m_i$ are non-negative
integers\footnote{$^{2}$}{It is understood that
we work in the space ${\cal V}$ of polynomial vectors, so
that $\A$, $\B$, etc. are defined on these. The space
${\cal V}$ is naturally graded by the degree of the
polynomials.} which satisfy the {\it resonance relation}
$$ \sum_{i=1}^n \ m_i \ \la_i = \la_r \ . \eqno(16) $$
Notice that this will be a polynomial of order $m = \sum_i
m_i$.

In the case of the identity matrix, $\la_i = 1$, and there
is no resonance relation with $m >1$. Thus, for $B=I$,
$\~F \in {\cal K}$ actually means that $\~F = 0$, and
therefore $\~f (y) = Ay$. This proves the Theorem. \hfill
$\odot$

{\tt Remark 5.} It should be stressed that the above proof
would also work if $B$ was not the identity, but any
matrix such that its semisimple part $B_s$ does not admit
resonances, as also discussed below. $\odot$

\bigskip\bigskip
{\bf 2. Convergence of the linearizing transformation.}
\bigskip

It is well known that in general the Poincar\'e procedure
is only formal, i.e. the series defining the coordinate
transformations (called in the following the {\it
normalizing transformation}, or NT for short) required to
take the system (6) into normal form is in general not
convergent.

Some special conditions which guarantee the convergence of
the NT are known; these deal either with the structure of
the spectrum of $A = (Df)(0)$ (e.g. the condition that
they belong to a Poincar\'e domain \ref{1-3}), either with some
symmetry property of $f$ (see the Bruno-Markhashov-Walcher
theory, \ref{16-19}; see also \ref{20,21}).

Here we just recall that if $A$ is real, its eigenvalues
$\sigma_1 , ... , \sigma_n $ belong to a Poincar\'e domain if and
only if $\epsilon {\rm Re} (\sigma_i ) > 0$ for all the $i$,
where the sign $\epsilon = \pm 1$ is the same for all $i$.
In this case, we are guaranteed of the convergence of the
Poincar\'e normalizing transformation \ref{1,2}.

{\bf Theorem 2.} {\it With the same notation and under the
same hypotheses as in theorem 1, the series of
near-identity changes of coordinates which takes $X$ and $Y$
into linear normal form is convergent in a neighbourhood of
the origin.}

{\tt Proof of Theorem 2.} In the case of the matrix $B=I$,
the eigenvalues are obviously in a Poincar\'e domain, and
the NT is therefore guaranteed to be convergent. Notice that
in general this NT would be not unique, being defined up to
elements in the kernel of the homological operator $\B$;
however, for $B=I$ this kernel is trivial, and the NT is
unique. The theorem 1 guarantees that this transformation
does also take $X$ into normal form, and thus that $X$ can
be linearized by means of a convergent change of coordinates.
\hfill $\odot$

{\tt Remark 6.} Similarly to what remarked above concerning
theorem 1, again the above proof would work (and theorem 2
apply) for more general symmetries: e.g., it would suffice
to require that the eigenvalues of $B$ belong to a
Poincar\'e domain (see also section 4). $\odot$

On the other hand, it is clear that the converse of Theorem 2 (and of
Theorem 1 as well) is also true: indeed, the VF $X$, once linearized,
obviously admits the dilation symmetry $S$, and, if the linearizing
transformation is convergent, there exists an analytic symmetry
$Y=p^i\pa_i$  with $(DY)(0)=I$, which is transformed into $S$ by the
transformation wich linearizes $X$. So, changing point of view,
and focusing on the convergence of
the normalizing transformation \ref{16-20}, the above arguments
can be summarized and completed in the following form:

{\bf Theorem 3.} {\it A vector field $X=f^i (x) \pa_i$ can be
linearized if and only if it admits a (possibly formal)
symmetry $Y=p^i (x) \pa_i$ with $(DY)(0)=B=I$; the normalizing
transformation which linearizes $X$ is convergent in a neighbourhood
of the origin if and only if this symmetry is analytic.}

Notice that theorem 3 includes the (trivial) case that the
vector field itself is such that $(DX)(0) = I$: the Poincar\'e
criterion (mentioned above) is in this case sufficient to
guarantee that the vector field can be linearized by a
convergent transformation: we can see this case as one in
which the symmetry requested by theorem 3 consists of the
vector field itself.

\bigskip\bigskip
\vfill\eject
{\bf 3. Families of vector fields.}
\bigskip

It should be stressed that the approach to the formal
linearization, and the proof that this is actually not only
formal, given above and based on symmetry properties of
the vector field does immediately extend to the case in
which we have a family of vector fields $X_\mu$ depending
smoothly on   real parameters $\mu \in R^m$ and such that
there is a family $Y_\mu$ of symmetry vector fields, i.e.
of vector fields such that $\[ X_\mu , Y_\mu \] = 0$,
provided the $Y_\mu$ have linear part $B (\mu ) \equiv (D
Y_\mu ) (0) = I$.

We are also assuming, in view of Remark 2 above, that the dynamics
is linearizable with respect to the same linear structure (i.e.,
independently of the parameter $\mu$). More in general, the arguments
given in Theorems 2 and 3 remain valid considering changes of coordinates
depending smoothly on $\mu$, and we are guaranteed to have a
$\mu$-dependent family of convergent normalizing transformations.

In this framework, (1) and (4) would now be
$$ X_\mu \ = \ \sum_{i=1}^n \ A_{ij} (\mu ) \ y_j \ {\pa ~
\over \pa y_i } \ , \eqno(1') $$
$$ y_i = \phi_i (x_1 , ... , x_n ; \mu ) \ . \eqno(4') $$
Under this, the linear dynamical system
$$ {\dot y_i} \ = \ A_{ij} (\mu ) \ y_j \eqno(5') $$
is changed into the nonlinear system
$$ {\dot x^i} \ = \ f^i_\mu (x) \ ; \eqno(6') $$
and thus in the new coordinates we have
$$ X_\mu \ = \ \sum_{i=1}^n \ f^i_\mu (x) \ {\pa ~ \over \pa
x_i} \ . \eqno(9') $$

We will not bore the reader by repeating any further our
previous discussion in the parameter-dependent case.

{\tt Remark 7.} We would like to stress that -- in the same
way as in the previous sections -- our discussion would also
apply to the case in which $B (\mu ) = (D Y_\mu )(0)$ is
not the identity, provided e.g. the eigenvalues $\la_i (\mu )$
of $B (\mu )$ belong to a Poincar\'e domain for all values
of $\mu$; see also the next section. $\odot$

{\tt Remark 8.} In order to avoid possible confusion, we would
like to briefly consider the case where we have a family of
vector fields $X_\mu$ admitting a common symmetry $Y$, i.e.
$[X_\mu , Y]=0$ $\forall \mu$.
If $B=(DY)(0)$ is the identity (or however does not admit
resonances, see also the next section), then according to our discussion
we would have a unique transformation which takes into normal form the
whole family of vector fields $X_\mu$ at once, and this can appear
surprising. Notice however that this is the case only if $B$
does not admit resonances, and in this case the $X_\mu$ would become linear
once transformed into NF.
Thus, if $B=I$, we know that any linear vector field $X_A = (Ax)^i
\pa_i$ commutes with $Y = (Bx)^i \pa_i$; if we change coordinates via a
near-identity transformation $x = \phi (y)$, then $Y = g^i(y) \pa_i$ is a
symmetry of all the vector fields $X_A = f^i(y) \pa_i$, which depend on
$(n \times n)$ parameters, i.e. the entries of the matrix $A$. $\odot$

\bigskip\bigskip
{\bf 4. Symmetries not having the identity as linear part.}
\bigskip

In some cases, determining a symmetry whose linear part is
{\it not} $B=I$ can also suffice to guarantee the -- formal
or convergent -- linearizability of the vector field, or at
least the possibility to considerably simplify it. This fact
was already pointed out in remarks 5-7 above, and we are now
going to discuss it a little further, in particular with
reference to the problem of convergence of the normalizing
transformation.

First of all, let us consider the case $B\not=I$ but
$\ker(\B)=\{0\}$, i.e. $B$ does not admit resonances (for
simplicity we assume $B$ semisimple): we
know that $f$ can then be linearized. If, in addition, the
symmetry $Y$ is analytic and the
matrix $B$ satisfies the ``condition $\omega$'' of Bruno [3],
then we are guaranteed that the normalizing transformation
which linearizes $f$ is convergent. This conclusion
follows from the Bruno theorem [3]: indeed the other
condition required by Bruno theorem ("condition A") is in
this case automatically satisfied, as the normal form is in fact
linear: $Y=(Bx)^i \pa_i$.

Let us recall briefly, for the sake of completeness, what are
the requirements for $B$ to satisfy ``condition $\omega$'':
we ask that $B$ is semisimple, and denote by $\la_i$ its
eigenvalues. Consider then the set of $Q=(q_1,\ldots , q_n)$
where $q_i$ are integers such that $q_i\ge -1$ and $(Q,\ \Lambda )
\neq 0$ (see [3] for full details), with
$$ (Q,\ \Lambda ) = q_1 \lambda_1 + \ldots  + q_n \lambda_n \
; \eqno(17) $$
let $\omega_k = \min |(Q,\ \Lambda)|$ on the $Q$ such that
$(Q, \ \Lambda ) \neq 0$ and $1 < \sum q_i < 2^k$. Then we say
that condition $\omega$ is satisfied if
$$ \sum_{k=1}^\infty 2^{-k} \ln \omega_k^{-1} \ < \ \infty \
. \eqno(18) $$

Let us then consider the case where again $B\not=I$, but its
eigenvalues belong to a Poincar\'e domain; in such a case
$f$ can be transformed, by means of a convergent series
of Poincar\'e transformations, into a very simple form
(even if possibly non-linear) i.e. can be brought to be in
$\ker ( \B )$, which is in this case finite dimensional.
Indeed, in this case we can take $Y$ into normal form, and
we are guaranteed the required normalizing transformation is
convergent (due to the Poincar\'e domain condition). In
doing this, $X$ is transformed as well, and in the new
coordinates $X = f^i(x) \pa_i$, with $f \in \ker (\B )$.

In the same vein, we can contemplate the case in which
$\ker (\A ) \cap \ker (\B ) = \{ 0 \}$ (this situation is
considered in example 3 below); in this case both $X$ and $Y$
are linearized when taken to the Joint Normal
Form -- as mentioned in \ref{9} -- but the (joint) normalizing
transformation is, without further assumptions, in general only
formal.

If $A$ satisfies condition $\omega$, then $f$ can be taken
into Normal Form -- i.e. in $\ker (\A )$ -- by a
convergent transformation, but we are not guaranteed in
general that the linearizing transformation is also convergent.

Notice, however, that this linearizing transformation is in fact
convergent in the particular case where $Y$ is linear, $Y=(B_{ij}  x_j )
\pa_i$. Indeed, the symmetry condition $[X,Y]=0$ becomes in  this case
$f\in\ker(\B)$, and it can be easily verified that this
condition is preserved by any transformation taking $f$  into Normal
Form; i.e., $f$ in $\ker(\A)$ implies  $f \in \ker (\A ) \cap \ker ( \B
)$ and therefore the Normal Form of $f$ is  necessarily linear.

If it is instead $B$ to satisfy condition $\omega$, then $f$ can be
taken to be in $\ker (\B )$ by a convergent transformation;
unless $B$ does not admit any resonance, this is obviously
not sufficient to ensure the convergent linearization of
$f$. Notice that in both these cases, ``condition A'' is
automatically satisfied.

The above considerations can be summarized in the following form.

{\bf Theorem 4}. {\it Let the vector fields $X=f^i \pa_i$ and $Y=g^i
\pa_i$ satisfy  $[X,Y]=0$, and assume the matrix $B=(Dg)(0)$ is
semisimple; then: {\tt a)} If $\ker(\B)=\{0\}$ and $B$ satisfies
condition $\omega$,  then $X$ and $Y$ can be linearized by a convergent
transformation; {\tt b}) If $\ker(\A)\cap\ker(\B)=\{0\}$, then $X$ and
$Y$ can be  linearized (possibly by means of a non-convergent
transformation), but  this transformation is convergent in the case $Y$
is linear, $Y=(Bx)^i \pa_i$.  Also, if $A=(Df)(0)$ (respectively
$B=(Dg)(0)$) is semisimple and satisfies condition  $\omega$, then there
is  a convergent transformation taking $X$ (respectively $Y$) into
Normal (not necessarily linear) Form.}

It is worth to emphasize the consequence of theorem 4 in the case of
families of vector fields. Indeed, for vector fields of the form (6')
one has that the eigenvalues of $A$ vary continously with $\mu$, so that
generically \ref{22}, for almost all values of $\mu$ the eigenvalues are
strongly nonresonant, and therefore, by Siegel therem the system can be
linearized by an analytic change of coordinates. Moreover, there is smooth (in
the sense of Whitney) dependence of the linearizing transformation on $\mu$,
when $\mu $ varies in the Cantor set to which correspond strongly nonresonant
frequencies. However, Siegel theorem does not give any information on
the behaviour of the system when the parameter belongs to the bad set
(the complementary of the above Cantor set).

Theorem 4 ensures that, provided there exists at least one $\mu$
dependent simmetry (with suitable properties), system (6') can be
linearized {\it for any value of the parameter}, and the linearizing
transformation depends smoothly on $\mu$, when $\mu $ belongs to an
interval, i.e. a regular set, and not a Cantor set.

We would like, to conclude this section, to point out that
it could happen that a symmetry with linear part the
identity cannot immediately be determined, but its
existence can be inferred from the presence of another
symmetry (and the linear space structure of the Lie algebra
of symmetries).

Thus, consider the case of a two-dimensional system $X = f^i(x)
\pa_i$ whose linear part $A$ is semisimple and has distinct
eigenvalues; assume that $X$ admits a
symmetry $Y = g^i(x) \pa_i$ whose linear part $B$ is semisimple
and not proportional to $A$. (Obviously one does
not have to ask explicitely all of these conditions: the
existence of distinct eigenvalues ensures $A$ is semisimple, and
then $[A,B]=0$ implies $B$ is semisimple as well.)
In this case, we are always able to easily find a symmetry
vector field $Z$ -- in particular, a linear combination
$Z=aX+bY$ -- whose linear part is the identity (it is for this
reason that in example 3 below we will have to consider a
three dimensional system).

The same argument is easily generalized to $n$-dimensional
systems $X$ having $n$ symmetries -- one of them being $X$
itself -- with independent linear part $B_i$, provided the
matrices $B_i$ are semisimple and span a nilpotent -- e.g.
abelian -- Lie algebra (for generalizations, not
needed in the present discussion, see \ref{13}).
Notice that, in particular, if $X$ is linearizable in the form (1) and
the $A^k$ are independent, these could be the vector fields $Y_{(k)}$
considered in (2).

\bigskip\bigskip
{\bf 5. Several symmetries having the identity as linear part.}
\bigskip

When we consider a fixed vector field $X$, there can be -- obviously --
different vector fields $Y$ which commute with it, i.e. which are
symmetries for $X$; the set of all vector fields which satisfy $[X,Y]=0$
is obviously a Lie algebra under the commutator; it is called the
symmetry algebra of $X$ and will be denoted by ${\cal G}$.

In particular, it can happen that several $Y_i \in {\cal G}$ satisfy
$(DY_i)(0) = I$. Notice that in general these do not form an algebra; in
particular, we know that $ \( D [Y_i , Y_j ] \) (0)  = \[ (DY_i )(0) ,
(DY_j )(0) \]$ (as it is immediately seen by expanding the $Y$ in Taylor
series around $0$), and therefore $[Y_i , Y_j ]$ has vanishing linear
part; thus, in particular, this linear part is not the identity. Notice
however that there is no reason for $[Y_i , Y_j ]$ to vanish in general.

According to our Theorem 1, we can choose each of the $Y_i$, and
simultaneously linearize $X$ and $Y_i$. It should be stressed that in
this way we do not, in general, linearize the other $Y_j $'s; more
precisely, not only we do not linearize them by the normalizing
transformation adapted to the pair $\( X , Y_i \)$, but in general we
are not able to linearize simultaneously different $Y$'s.

This fact has to do with the problem of simultaneously reducing to Joint
Normal Form a general algebra of vector fields \ref{10,23}. We
mention just the case of interest here, i.e. of algebras of vector
fields having semisimple linear part (see \ref{13} for a more general
discussion): in this case, one can prove that a Joint Normal Form is
possible only if the algebra is nilpotent, which includes in particular
the case of abelian algebras \ref{10}.

It should be mentioned that the case of general -- or even just solvable
-- algebras appears to be extremely hard; indeed, it is related to the
problem of simultaneous reduction to Jordan form of an algebra of
matrices (the linear parts of the vector fields in question); or, this
problem is not solved, neither it is known under which conditions it can
or cannot be solved \ref{24}.

{\tt Remark 9.} This is also maybe an appropriate point to remark
that in Theorem 1 we could only guarantee the existence of a
transformation linearizing both $X$ and $Y$; however, it is
quite clear -- {\it a fortiori} in the light of the above
considerations -- that we could also have a transformation which
linearizes $X$ without linearizing $Y$ (or viceversa).
Similarly, in general the transformation which linearizes
$X$ and $Y$ will not be unique. $\odot$

\bigskip\bigskip
{\bf 6. Examples.}
\bigskip

We will now consider some very simple example. Examples 1-3 show how
we can apply our results to guarantee that a concrete nonlinear system
can be linearized, without actually performing the Poincar\'e
normalization; example 4 deals with the situation discussed in sect.5,
and the somewhat artificial example 5 (see below for its construction)
wants to show how the method can deal with ``seriously wrong'' initial
coordinates.

\bigskip

{\tt Example 1.}  Consider the dynamical system in $R^2$ given by
$$ \eqalign{ {\dot x} \ =& \ x \cr {\dot y} \ =& \ 3y - x^2 \ . \cr}
\eqno(19) $$
The linear part of this is given by the matrix
$$ A = \pmatrix{1&0\cr0&3} \ , \eqno(20) $$ and the general Normal Form
corresponding to such a linear part is given by
$$ \eqalign{ {\dot x} \ =& \ x \cr {\dot y} \ =& \ 3y +
\alpha x^3  \cr} \eqno(21) $$ with $\alpha$ an arbitrary real constant.

By explicit (standard) computations, one can check that the symmetry
algebra of (19) is spanned by the vector fields $$ Z_1 = x \pa_x + 2 x^2
\pa_y \qquad ; \qquad Z_2 = (y-x^2 ) \pa_y \eqno(22) $$ (in particular,
$X = Z_1 + 3 Z_2$ corresponds to (19) itself); if we choose $Y = Z_1 +
Z_2$, i.e. $$ Y = x \pa_x + (y + x^2 ) \pa_y \ , \eqno(23) $$ we have
found a symmetry vector field $Y$, whose linearization is indeed
$(DY)(0) = I$. This guarantees that our system can be linearized, and
actually provides the linearizing transformation as well.

In the above example, we had a very simple situation, as no parameter is
appearing in the vector field, and the eigenvalues belong to a Poincar\'e
domain, so that the convergence of the normalizing transformation is
guaranteed. Thus, the only nontrivial result is that the term $\alpha
x^3$ in (21) does actually disappear from the normalized form. In the
following examples, we consider more complicate cases.
\bigskip

{\tt Example 2.} Let us consider again a two-dimensional system, i.e.
$$ \eqalign{  {\dot x} =& ~~x +x^4y \cr  {\dot y} =& -2y -x^3 y^2 \ .
\cr} \eqno(24) $$

The linear part of this is given by the matrix
$$ A = \pmatrix{1&0\cr0&-2} \ , \eqno(25) $$ and the general Normal Form
corresponding to such a linear part is given by
$$ \eqalign{  {\dot x} =& ~~x +x \, \phi_1 (x^2 y) \cr
{\dot y} =& -2y +y \, \phi_2 (x^2 y) \ . \cr} \eqno(26) $$
where $\phi_i$ are two arbitrary (smooth) functions.

A symmetry vector field of this is given by  $$ Y = (x + 4 x^4 y) \pa_x
+ (y - 4 x^3 y^2 ) \pa_y \ ; \eqno(27) $$ this has linear part given by
$B=I$, and thus we conclude the system (24) can be reduced to its linear
part by a convergent normalizing transformation.

Notice that in this case $\ker (\A )$ is infinite dimensional, and the
eigenvalues do not belong to a Poincar\'e domain.
\bigskip

{\tt Example 3.} We consider now a (fourth order) system in
$R^3$:
$$ \eqalign{  {\dot x} =& ~~x + a_1 x^3 y   + b_1 x y^2 z \cr  {\dot y} =& -3y +
a_2 x^2 y^2 + b_2 y^3 z \cr {\dot z} =& ~9z + a_3 y^2 z^2 + b_3 y^2 z^2 \ ,
\cr}
\eqno(28) $$  where $a_i,\ b_i$ are arbitrary constants; thus,
$$ A = \pmatrix{1&~0&0\cr0&-3&0\cr0&~0&9\cr} \ . \eqno(29) $$

A (linear) symmetry for this DS is given by
$$ Y = x \pa_x - 2 y \pa_y + 4 z \pa_z \ , \eqno(30) $$ and we have
$$ B = \pmatrix{1&0&0\cr0&-2&0\cr0&0&4\cr} \ . \eqno(31) $$
In this case, both $\ker (\A )$ and $\ker (\B )$ are infinite
dimensional, but their intersection is just $\{ 0 \}$. Notice also that
their eigenvalues (both for $A$ and $B$) are not in a Poincar\'e domain.

According to the arguments in Sect. 4, we can conclude that $X$ can be
linearized by a convergent transformation.

\bigskip

{\tt Example 4.} In this example we merely want to illustrate
the discussion of sect.5, and consider a
case (strongly related to one already considered in \ref{5}, see example
6 there -- notice however that here we
make different hypotheses concerning the linear part) in
which we have several noncommuting symmetries $Y$ of a given
vector field $X$, such that $(DY)(0)=I$. To avoid unnecessary
complications, we consider
$X$ to be already in linear form; it is of course possible to rephrase
the example by setting $X$ to be nonlinear, by means of any suitable
change of coordinates.

Consider the linear vector field
$$ X \ = \ x {\pa ~ \over \pa y} - y {\pa~ \over \pa x} \ ; \eqno(32)$$
it is immediate to check that any vector field of the form
$$ Y \ = \ f(r^2) X + g(r^2) Z \eqno(33) $$ is a symmetry of $X$, where
$$ Z = x {\pa \over \pa x} + y {\pa \over \pa y} \eqno(34) $$
and $r^2 = x^2 + y^2 $.

In particular, if we choose $f,g$ such that
$$ f(0) = 0 \qquad ; \qquad g(0) = 1 \eqno(35) $$ we have a symmetry
vector field with linear part the identity.

We can rewrite such vector fields as
$$ Y_{f,h} \ = \ f(r^2 ) X + \( 1 + h(r^2) \) Z \ , \eqno(36) $$ where
it is understood that both $f$ and $h$ vanish in zero.

One can readily check that
$$ \eqalign{
\[ Y_{f,h} , Y_{\phi , \eta } \] =&  2 r^2 \[ (1+h) \phi' - (1+\eta )
f' \] X + \cr
~ & + 2 r^2 \[ (1+h) \eta' - (1+\eta ) h' \] Z \ \ . \cr} \eqno(37) $$
and thus that the $Y_{f,h}$ do not commute among themselves. More
precisely, $Y_{\phi, \eta} $ commutes with $Y_{f,h}$ if and only if
$$ \cases{ \phi = c_1 f + c_2 & \cr 1 + \eta = c_1 (1 + h) & \ , \cr}
$$
with $c_1 , c_2$ two arbitrary real numbers;
it should be noticed that when we linearize $Y_{f,h}$, which reduces
then to $Z$, these commuting fields $Y_{\phi , \eta }$ do also reduce to
$Z$, as such fields also have the identity as linear part.

>From this short discussion, we can draw several conclusions. First of
all, $X$ admits a symmetry with linear part the identity; however, $X$
is already linear, so we should not enquire about it being linearizable.

As for the $Y_{f,h}$, these all admit the linear vector field
$X$ as symmetry, but $(DX)(0) \not= I$; thus, the first part of
theorem 1 cannot guarantee the linearizability of $Y$. On the
other side, the second part of the same theorem does guarantee
that it is possible to take $Y$ into linear form; this is not
really a surprise, as the linear part $Y_0$ of $Y$ is associated
to identity matrix $I$ (indeed $Y_0 (x) = I x$), and the
eigenvalues of this are not resonant in the Poincar\'e sense,
and moreover belong to a Poincar\'e domain
\ref{1}, so that the same result could be obtained by classical
Poincar\'e theory.

Finally, for what concerns linearizing different $Y$ at the
same time,  the above computation for the commutator shows that
in general -- i.e. unless (37) vanishes -- we are not able to
simultaneously linearize different $Y_{f,h}$.

{\tt Remark 10.} The vector fields $Y_{f,h}$ were also considered in
example 6 of \ref{5}, and found to be non-linearizable. In order to avoid
possible misunderstandings, it should be stressed that the hypotheses made
on the linear parts -- and in particular on the role of the rotation component
of the vector field in its linearization -- are different here and there. It
should also be mentioned that there we
considered global linearization, while here we are in the (perturbative)
framework of normal forms theory, and we consider only linearization in
a neighbourhood $U$ of the origin. This point is further discussed
in the Appendix. $\odot$

\bigskip


{\tt Example 5.} As a final example, we consider an apparently
hopelessly complicate system (see below for how it was
generated), i.e.
$$ \eqalign{ {\dot x} = f_1 (x,y,z) =  & \[ \a x  - y \] - x^2  - \[ 3 x y^2 + 2
\a y^3 \] \cr & - 6 \[ x^3 y + \a x^2 y^2 \] - 3 \[ x^5 + 2 \a x^4 y + y^5 \] -
\[ 2 \a x^6 + 15 x^2 y^4 \] \cr & - 30 \[ x^4 y^3 + x^6 y^2 \] - 3 \[ x^{10} + 5
x^8 y \] \cr {\dot y} = f_2 (x,y,z) = & \[ x + \a y \]
 - \[ \a x^2 - 2 x y \]
 + \[ 2 x^3 + y^3 \] \cr &
 + \[ 9 x^2 y^2 + 4 \a x y^3 \]
 + \[ 15 x^4 y  + 12 \a x^3 y^2 \]
 + \[ 7 x^6 + 12 \a x^5 y + 6 x y^5 \] \cr &
 + \[ 4 \a x^7 + 30 x^3 y^4 \]
 + 60 x^5 y^3
 + 60 x^7 y^2
 + 6 \[ 5 x^9 y + x^{11} \] \cr {\dot z} = f_3 (x,y,z) = & \b z
 + \[ 2 x y + \( 2 \a - \b \) y^2 \] \cr &
 + \[ 2 x^3 + 2 \( 2 \a - \b \) x^2 y + 3 x y^2 + \a y^3
    + \( 2 \a - \b \)  y^3 \] \cr & + \[ \(2 \a - \b \) x^4 - 3 \a x^2 y^2 + 6 x
y^3 + 2 y^4 \]
  + \[ 6 x^3 y^2 + 8 x^2 y^3 + 3 y^5 \] \cr &
  + \[ 12 x^4 y^2 + 27 x^2 y^4 + 12 \a x y^5 \]
  + 9 \[ 5 x^4 y^3 + 4 \a x^3 y^4 \] \cr &
   + \[ 2 x^8 + 8 x^6 y + 21 x^6 y^2  + 36 \a x^5 y^3
      + 18 x y^7 \]
 + \[ 12 \a x^7 y^2 + 90 x^3 y^6 \] \cr &
 + 180 x^5 y^5
 + 180 x^7 y^4
 +  90 x^9 y^3
 +  18 x^{11} y^2
\ .  \cr}
\eqno(38) $$

In this case the linear part is given by
$$ {\dot {\bf x}} = A {\bf x} \eqno(39) $$ with ${\bf x} = (x,y,z)$ and $A$ the
matrix
$$ A = \pmatrix{\a & -1 & 0 \cr 1 & \a & 0 \cr 0 & 0 & \b
\cr} \ .  \eqno(40) $$

Despite the very complicate form of (38), one can check explicitely that $X =
f_i ({\bf x}) ( \pa / \pa x_i )$ commutes with  $Y = g_i ({\bf x}) ( \pa / \pa
x_i )$  (here and in the following of this example, $(x_1 , x_2 , x_3 )  =
(x,y,z)$ ), where
$$ \eqalign{ g_1 ({\bf x}) = & x - 2 y^3 - 6 x^2 y^2
  - 6 x^4 y - 2 x^6 \cr g_2 ({\bf x}) = & y - x^2 + 4 x y^3 + 12 x^3 y^2 +
4 x^7
  + 12 x^5 y \cr g_3 ({\bf x}) = & z + y^2 + 2 x^2 y + 2 y^3 + x^4
  - 3 x^2 y^2 + 12 x y^5 \cr &
  + 36 x^3 y^4 + 36 x^5 y^3 + 12 x^7 y^2 \ .
\cr} \eqno(41) $$

The linear part of this vector field corresponds just to the identity matrix,
and thus our theorem guarantees immediately that $X$ can be reduced to its
linear part; i.e. in Normal Form coordinates we have
$$ X = \( A_{ij} y_j \) {\pa \over \pa y_i }\ . \eqno(42) $$

Actually, the system (38) was obtained from (42) by the change of coordinates
$$ \eqalign{ y_1 = & x_1 + \( x_1^2 + x_2 \)^3 \cr y_2 = & x_2 + x_1^2 \cr y_3 =
& x_3 - x_2^3 - \( x_1^2 + x_2 \)^2 \ . \cr}
\eqno(43) $$ and $Y$ is, in the $y$ coordinates, nothing else than the dilation
vector field, i.e.
$$ Y = y_i {\pa \over \pa y_i } \ . \eqno(44) $$ Similarly, we could have
considered the vector field
$$ Z = \( (A^2)_{ij} y_j \) {\pa \over \pa y_j} \eqno(45) $$ which depends on
$\a$ and $\b$, and which obviously commutes with $X$ (and with $Y$).

We stress that now we have a two parameters family of systems (42), and that as
$\a , \b $ are varied, this goes through resonances, and the eigenvalues
$\la_\pm = \a \pm i$ and $\la_0 = \b $ can be in a Poincar\'e domain or
otherwise. However, our symmetry method is not sensitive to these facts, and
works for all values of the parameters.

\bigskip\bigskip
{\bf 7. Some further remarks.}
\bigskip

{\tt Remark 11.} In the previous example 1, we have been able to
identify {\it all} the symmetries of our system. We would like
to stress, however, that the only important fact from our point
of view is that we are able to determine {\it one}  symmetry
with the required linear part: this is a much simpler task, and
this is what has been done in the other examples. As it is
generally the case with symmetry methods for differential
equations, it is the possibility to obtain relevant informations
from the knowledge of {\it one} symmetry (or a symmetry
subalgebra), without the need to know the full -- in general,
infinite dimensional -- symmetry algebra of the dynamical system
under study, which makes our method applicable. $\odot$

{\tt Remark 12.} In general, one could try to determine
perturbatively the functions $p^i (x)$, by expanding them in
homogeneous terms as $p^i (x) =
\sum_{m=1}^\infty p^i_m (x)$, where $p^i_m (a x) = a^m p^i_m (x)$, and
solving the determining equations order by order \ref{25}; in
particular, one should require $p^i (x) = x_i$. It should also be
mentioned that if in this way we determine a solution (or a solution
exists) only up to some order $k$, we can guarantee that the system can
be linearized up to terms of order $k$ \ref{25}.

This information, although more limited than a full linearization
property, can equally be of great utility: first, because in
actual computations one does in most cases consider a truncation
at some (high) order $k$; and second, because if we combine such
a result with the analysis of resonances, in order to guarantee
the full linearizability of the system it suffices to guarantee
the existence of a symmetry $Y$ with linear part the identity up
to the order $k$ of the highest order resonance of the system
(e.g., in example 1 above one would only had to go up to order
three). Notice that we are guaranteed that this order is finite
when the eigenvalues belong to a Poincar\'e domain.  $\odot$

{\tt Remark 13.} It can be helpful to see theorem 1 from a
slightly different perspective than the one used here in
section 1 (we use freely the notation employed there); this
discussion does actually repeat points already mentioned, and
is thus also a way to summarize our argument.
As $(DY)(0) = I$, we know \ref{1} that $Y$ is
biholomorphically equivalent to its linear part in a
neighbourhood $U$ of the origin (this only depends on the
fact that the eigenvalues of $(DY)(0)$ belong to a Poincar\'e
domain and there are no resonances, so that it would extend to
more general linear parts $(DY)(0) = B$). When we apply the
normalizing transformation we have, denoting by $y$ the
``new'' coordinates, $Y = y^i \pa / \pa y^i$, and $X =
{\~f}^i (y) \pa / \pa y^i$; however, the relation $[X,Y] = 0 $
is independent of the coordinate representation of $X$ and $Y$,
and thus in the neighbourhood $U$ in which the
normalizing transformation is not only formal (actually, as
mentioned above, is biholomorphic) we necessarily have
${\~f}^i (y) = A^i_j y^j$ with $A$ a matrix, and actually $A =
(DX)(0)$. $\odot$

\def\Ck{{{\cal C}^k}}
{\tt Remark 14.} The previous remark also shows what kind of results
should be expected when we deal with $\Ck $  functions rather than with
formal power series: if $Y$ is a $\Ck $ vector field with e.g. $(DY)(0) =
I$, we can put it into normal form -- i.e. linearize it -- by a $ {\cal
C}^{k-1}$ transformation; the rest of the argument follows as before, and
thus we conclude that if $X,Y$ are commuting $\Ck$ vector fields, with
the same hypotheses as above concerning their linear parts, then they can
be simultaneously linearized by a ${\cal C}^{k-1}$ normalizing
transformation. We thank N.N. Nekhoroshev for this observation. $\odot$

{\tt Remark 15.} Finally, we would like to point out that the
present approach is related to the study of integrability
conducted by Marmo and collaborators, see e.g. \ref{26,27}; the
use of symmetries to study the linearizability of a dynamical
system has been considered, in a non-perturbative approach
(related to the  general theory of symmetry methods for
differential equations \ref{28-31}), in \ref{5}. $\odot$


{\bf Acknowledgements.}

We would like to thank the referees of a previous version of this
paper, who have forced us to clarify several points.

One of the authors (G.G.) would like to thank the Mathematische
Forschunginstitut Oberwolfach  and the I.H.E.S. (Bures sur Yvette) for
hospitality during part of this work. The stay in M.F.O. was supported
by the {\it Volkswagen Stiftung} under the RiP program.

\eject

{\bf APPENDIX: On perturbative and non-perturbative
linearization.}

\bigskip

As already pointed out in the Introduction and in the body of the
paper, in the present work we discuss the same problem as in \ref{5},
i.e. the relation between symmetry and linearizability for Dynamical
Systems; the main difference between the two approaches being that there
we used a non-perturbative and global approach, while here we remain in
the framework of perturbation theory. However, it is quite clear that
(strong) relations have to be present between the two approaches, and
this appendix is devoted to the discussion of these.

First of all, in the perturbative approach one should distinguish
between formal linearization and actual one, i.e. -- in the present
setting based on normal forms -- between the case in which the
normalizing transformation providing the linearization of the system is
purely formal, and the case where it converges.

It is quite clear that in the case of purely formal
linearization, we should expect in general that
the system is not linearizable when we consider global transformations,
i.e. in the sense adopted in \ref{5}.

The second point is that, even when the normalizing transformation  is
convergent, this transformation is in general by no means global: i.e.,
it is defined only on some neighbourhood $U$ of the origin. Thus,
even in this case, it is not surprising if we get results which do not
agree with those obtained by the approach proposed in \ref{5}.

It should also be pointed out that it is possible that a system can be
linearized, globally or only in a neighbourhood of the origin, but {\it
not} by a near-identity transformation: in this case, it cannot be
linearized by the Normal Forms approach.
An example of this situation is provided by
$$ {\dot x_i} \, = \, \sum_{j=1}^n \ A_{ij} \, {\rho^2 \, x_j \over
\rho^2 + x_i^2} \eqno(A1) $$
(with $\rho^2 = \sum_{i=1}^n x_i^2$), which
is nothing else than  $$ {\dot y_i} \, = \, A_{ij} \, y_j \eqno(A2) $$
with the change of coordinates $y_i = \rho^2 x_i$.

Let us now come back to the differences between global and local
linearization; a clarifying example in this respect is provided by a
nonlinear oscillator described by (here $r^2 = x^2 + y^2$)
$$ \eqalign{
{\dot x} \ = & \ - r^2 y -  (r^2 - 1) x \cr
{\dot y} \ = & \ + r^2 x -  (r^2 - 1) y \ ; \cr} \eqno(A3) $$
this can be solved (e.g. passing to polar coordinates), and it
is clear by its behaviour that it cannot be linearized in global sense,
according to \ref{5}, by the same arguments used in \ref{5}. On the
other side, its linear part is just the identity, and thus by
the standard results in normal forms theory discussed here,
it can be linearized by a convergent transformation (in fact, it
is biholomorphically equivalent to its linear part in a
neighbourhood $U$ of the origin \ref{1}), but it is
obvious that such a neighbourhood $U$ cannot include the limit
cycle.

Another example, which we want to discuss at some length in the
following, is provided by the vector fields $Y_{f,h}$ considered in
example 4 of section 6.

Thus, we have that -- not surprisingly -- global and local
linearization are quite different.
However, the discussion conducted in \ref{5} could be restricted to a
neighbourhood $U$ of the origin; in this case, the results obtained by
the two approaches should be -- when the normalizing transformation is
convergent -- equivalent.

We are thus presenting here a short discussion of the restriction of
the approach of \ref{5} to such a neighbourhood; we will, for the sake
of clarity, discuss systems as those encountered in example 4 above
(to avoid confusion with the notations used in this example and in (A3)
as well, we are introducing here new and independent notations).
Precisely, we want to discuss the problem of existence of a diffeomorphism
\footnote{$^3$}{In the present discussion, a diffeomorphism
will be meant to be of class ${\cal C}^1$, and not necessarily ${\cal
C}^\infty$.} $\Phi : U \to U$ conjugating a vector field $Y$, such that
$Y(0) = 0$, to the dilation vector field
$$ \~Y \ = \ \eta_1 { \pa ~~\over \pa
\eta_1} \, + \, \eta_2 { \pa ~~\over \pa \eta_2} \ \ . \eqno(A4) $$
As $Y(0) = 0$ and $\~Y (0) = 0$, we also require that $\Phi (0) = 0$.

\def\T{{\rm T}}
First of all, we notice that the eigenvalues of $A = (DY) (0)$ are
invariants under any such $\Phi$ (of class at least ${\cal C}^1$);
indeed, $Y$ induces its lift $Y^T$ on $\T U$, given (with $v^i $
spanning a basis in $\T_x U$) by
$$ Y^T \ = \ Y^i {\pa ~ \over \pa x^i} \ + \ {\pa Y^i \over \pa x^j} \,
v^j {\pa ~\over \pa v^i} \eqno(A5) $$
and, as $Y(0)=0$, we have
$$ Y^T (0) \ = \ {\pa Y^i \over \pa x^j} (0) \,
v^j {\pa ~\over \pa v^i} \ \equiv \ A^i_j v^j {\pa ~\over \pa v^i} \ .
\eqno(A6) $$

Also, $\Phi : U \to U$ induces a diffeomorphism $\Phi^T : \T U \to \T
U$; thus for conjugated fields $Y , \~Y$, $Y^T$ and $\~Y^T$ are
conjugated in the origin by $\Phi^T$. We conclude that all the
algebraic invariants of $(DY)(0)$ and of $(D\~Y)(0)$ coincide.
Thus we have that a {\it necessary} condition for $Y$ and $\~Y$ to be
${\cal C}^1$-conjugated is that the matrices $A = (DY)(0)$ and $\~A =
(D \~Y )(0)$ are conjugated.

In the case of
$$ Y \ \equiv \ Y_{f,h} \ = \ [1 + h(r^2 ) ] \, Z \ + \ f(r^2) \, X
\eqno(A7) $$
we have thus that a necessary condition for $Y$ to be
conjugated to $\~Y$ is that $A = (DY)(0)$ is conjugated to the identity
matrix. But in this case $A$ is just the identity matrix, as
$f(0)=h(0)=0$.

If we consider the more general family of vector fields (see example 6
of \ref{5})
$$ Y_{\a , \b} \ = \ \a (r^2 ) Z + \b (r^2 ) X \eqno(A8) $$
we have that
$$ A \ = \ \pmatrix{ \a (0) & - \b (0) \cr \b (0 ) & \a (0) \cr}
\eqno(A9) $$
and thus we have as necessary condition that the
determinant and trace of $A$ are equal to that of the identity matrix;
these two conditions together mean that $a(0) = 1$, $b(0)= 0$, i.e. $A = I$
(obviously the identity matrix can only be conjugated to itself). This
condition is verified in (A7), but not in the case considered in \ref{5}.

Although a necessary condition can be useful to ensure
linearizability, it would be preferable to have a necessary and
sufficient condition; this is provided by existence of two solutions
$\eta_1$ and $\eta_2$, such that $\eta_1 (0) = \eta_2 (0) =0$ and
functionally independent on $U$, to
$$ L_Y \, \eta \  = \ \eta  \eqno(A10) $$
(where, as in the following, $L_Y$ is the Lie derivative
along the vector field $Y$).
We could of course attempt a solution of this by series expansion: this
would be essentially equivalent to the Poincar\'e method.

In the case of $Y = Y_{f,h}$ this equation reads
$$ L_Z \eta + f(r^2 ) L_X \eta + h(r^2) L_Z \eta \ = \ \eta \ . \eqno(A11) $$

In the general case, it can also be appropriate to observe that if $Y$
is conjugated in $U\subseteq R^n$ (with $\{ O \} \in U$) to the
linear field
$$ Y_A \  = \ A_i^j \eta_j {\pa \over \pa \eta_i} \ , \eqno(A12) $$
then there will exist $n$ functionally independent solutions to
$$ L_Y \eta_i \ = \ A_i^j \eta_j \ . \eqno(A13) $$

It should be stressed that in this way we can show that if $Y$ is
locally conjugated to the dilation field $\~Y$, then necessarily $(DY)(0)
= \pm I$; this is the converse to the statement, well known in Normal
Form theory, that if $(DY)(0) = \pm I$ then $Y$ is locally conjugated to
the dilation field.

It should be stressed also, for completeness of discussion, that a
vector field can be somehow correlated to the dilation field, e.g.
being proportional to it, without being conjugated to it. Indeed,
consider the vector field
$$ Y \ = \ {1 \over 3} \, x {d \over dx} \eqno(A14) $$
and let us look for solutions to (A10). This yields $(1/3) x d\eta /dx =
\eta$ and thus $\eta (x) = x^3$; thus $\eta$ is analytic, but it does
not define a diffeomorphism (its inverse $x = \eta^{1/3}$ is singular
in the origin).

Finally, we would like to mention that, if we place ourselves in the
point of view adopted in this Appendix, the discussion presented in the
main body of the paper can be reinterpreted as a discussion of
perturbative techniques, i.e. of the use of Poincar\'e theory, to solve
(A11).

~\bigskip\bigskip\bigskip
~\bigskip\bigskip\bigskip

\parskip=4pt

{\bf References}
\bigskip
\parindent=15pt

\item{[1]} V.I. Arnold, ``Geometrical methods in the theory of differential
equations'', Springer, Berlin 1988

\item{[2]} V.I. Arnold and Yu.S. Il'yashenko, ``Ordinary differential
equations''; in {\it Encyclopaedia of Mathematical Sciences - vol. I,
Dynamical Systems I}, (D.V.
Anosov and V.I. Arnold eds.), p. 1-148, Springer, Berlin 1988

\item{[3]} A.D. Bruno, ``Local methods in nonlinear differential equations'',
Springer, Berlin 1989

\item{[4]} F. Verhulst, ``Nonlinear differential equations and dynamical
systems'', Springer, Berlin 1990, 1996$^2$

\item{[5]} G. Gaeta and G. Marmo,  {\it Journ. Phys. A: Math. Gen.} {\bf 29}
(1996), 5035

\item{[6]} G. R. Belitsky, {\it Russ. Math. Surveys} {\bf 33} (1978) 107

\item{[7]} C. Elphick, E. Tirapegui, M.E. Brachet, P. Coullet, and G. Iooss,
{\it Physica D} {\bf 29} (1987) 95

\item{[8]} G. Iooss and M. Adelmeyer ``Topics in bifurcation theory and
applications, World Scientific, Singapore 1992

\item{[9]} G. Cicogna and G. Gaeta, {\it Journ. Phys. A: Math. Gen.} {\bf 27}
(1994) 461

\item{[10]} G. Cicogna and G. Gaeta, {\it Journ. Phys. A: Math. Gen.} {\bf 27}
(1994) 7115

\item{[11]} G. Benettin, L. Galgani and A. Giorgilli, {\it Nuovo Cimento B}
{\bf 79} (1984), 201

\item{[12]} Yu. A. Mitropolsky and A.K. Lopatin, ``Nonlinear mechanics, groups
and symmetry'', Kluwer, Dordrecht 1995

\item{[13]} G. Cicogna and G. Gaeta, ``Symmetry
and perturbation theory in nonlinear dynamics'', preprint
IHES P/97/12

\item{[14]} M.W. Hirsch and S. Smale ``Differential equations, dynamical
systems, and linear algebra'', Academic Press, London 1974

\item{[15]} A.N. Kolmogorov and S. Fomine, ``Elements of the Theory of
Functions'', Pergamon

\item{[16]} L.M. Markhashov, {\it J. Appl. Math. Mech.} {\bf 38} (1974), 788

\item{[17]} S. Walcher, {\it Math. Ann.} {\bf 291} (1991), 293

\item{[18]} A.D. Bruno, {\it Selecta Mathematica} {\bf 12} (1993), 13

\item{[19]} A.D. Bruno and S. Walcher, {\it J. Math. Anal. Appl.} {\bf 183}
(1994), 571

\item{[20]} G. Cicogna, {\it J. Phys. A: Math. Gen.} {\bf 28}
(1995), L179; {\it J. Math. Anal. Appl.} {\bf 199} (1996), 243;
Preprint {\tt mp\_arc 97-159}, 1997 (to appear in {\it J. Phys. A: Math. Gen.}

\item{[21]} H. Ito, {\it Comm. Math. Helv.} {\bf 64} (1989), 412; {\it Math.
Ann.} {\bf 292} (1992), 411

\item{[22]} V.G. Sprindzuk: ``Metric Theory of Diophantine Approximation'',
Winston and Sons, Washington 1979

\item{[23]} D. Arnal, M. Ben Ammar and G. Pinczon, {\it Lett. Math. Phys.}
{\bf 8} (1994), 467

\item{[24]} A.A. Kirillov, ``Elements of the Theory of Representations'',
Springer, Berlin 1976

\item{[25]} G. Cicogna and G. Gaeta, {\it Nuovo Cimento B} {\bf 109} (1994),
989

\item{[26]} S. De Filippo, G. Vilasi, G. Marmo and M. Salerno,  {\it Nuovo
Cimento B} {\bf 83} (1984), 97

\item{[27]} G. Marmo, in ``Proceedings of the third seminar on Group
Theoretical Methods in Physics'' (Yurmala, May 22-24 1985), Nauka,
Moscow 1986

\item{[28]} P.J. Olver, ``Applications of Lie groups to differential
equations'', Berlin, Springer 1986

\item{[29]} L.V. Ovsjannikov, ``Group properties of differential
equations'' (Novosibirsk,  1962; English transl. by G. Bluman, 1967);
``Group analysis of differential equations'', Academic Press, New York
1982

\item{[30]} G.W. Bluman and S. Kumei, ``Symmetries and differential
equations'', Springer, Berlin 1989

\item{[31]} G. Gaeta, ``Nonlinear symmetry and nonlinear equations'',
Kluwer, Dordrecht 1994

\bye